\documentstyle[preprint,aps]{revtex}
\begin{document}
\input{psfig.sty}
\def \gam {\frac{ N_f N_cg^2_{\pi q\bar q}}{8\pi} }
\def \gamm {N_f N_cg^2_{\pi q\bar q}/(8\pi) }
\def \be {\begin{equation}}
\def \ba {\begin{eqnarray}}
\def \ee {\end{equation}}
\def \ea {\end{eqnarray}}
\def \gap {{\rm gap}}
\def \gapp {{\rm \overline{gap}}}
\def \gappp {{\rm \overline{\overline{gap}}}}
\def \im {{\rm Im}}
\def \re {{\rm Re}}
\def \Tr {{\rm Tr}}
\def \P {$0^{-+}$}
\def \S {$0^{++}$}
\def \uu {$u\bar u$}
\def \dd {$d\bar d$}
\def \ss {$s\bar s$}
\def \qq {$q\bar q$}
\def \qqq {$qqq$}
\title{
Flavor Symmetry as a Spontaneously Broken Discrete Permutation Symmetry 
Embedded in Color }
\author{Nils A. T\"ornqvist}
\address{Physics Department, 
POB 9, FIN--00014, University of Helsinki, Finland}
\date{August 21, 1999}                 
\maketitle
\begin{abstract}
A new mechanism for breaking an internal symmetry spontaneously is discussed,
 which is intermediate between the Nambu-Goldstone and Wigner modes of 
symmetry breaking. Here the $q\bar q$ sea takes the role of the vacuum 
of the Nambu-Goldstone case. Flavor symmetry becomes a discrete permutation 
symmetry of the valence quarks with respect to the sea quarks, which can be 
spontaneously broken without generation of massless Goldstone bosons.
\\
\noindent Pacs numbers:12.39.Ki, 11.30.Hv, 11.30.Qc, 12.15.Ff
\vskip 0.90cm 
\end{abstract}

It is a well known fact that most hadrons are built from $q\bar q $ or $qqq$
valence quarks together with a $q\bar q$ and gluon sea.
This two-component picture is crucial for my mechanism, 
to be discussed below, where I shall give the finite $q\bar q$  
sea a prominent 
role in the symmetry breaking, which usually is given to the vacuum, when a 
symmetry is spontaneously broken. I consider flavor symmetry and take 
$N_f=N_c=3$ in the discussion, although the same arguments should hold for 
any number of flavors and colors, and might be applied to other symmetries as 
well (like chiral symmetry).

If we undress a hadron from its soft confined gluons the \qq\ valence quarks 
of a meson can be thought of as a degenerate nonet in color (and the \qqq\ 
valence quarks of a baryon as a 
${\bf 3\otimes 3\otimes 3 =1\oplus 8\oplus 8\oplus 10}$ multiplet). 
Likewise the undressed \qq\  sea is composed of a nonet and higher 
representations in color. After dressing with  soft gluons, the hadrons 
become singlets\footnote{A mechanism by which one can understand this is 
to assume all gluonic transitions, 
$q_i\bar q_jglue \to q_{i^\prime }\bar q_{j^\prime }glue$, 
between the N degenerate states to be equal $H_{ij,i' j' }=const$. 
After diagonalization this gives 0 for all other transitions than singlet 
to singlet which is $N\cdot const$, i.e., all other states except 
the singlet decouple.} in color.

Now chose a particular global reference frame in color, in which the sea 
becomes diagonal, such that it can be composed of diagonal Gell-Mann matrices 
$\lambda_i$: 
$S(q\bar q)=\epsilon_0\lambda_0+\epsilon_3\lambda_3+\epsilon_8\lambda_8=diag(x,y,z)$. 
This picks out a special direction and ordering in color space. One can still 
permute the $x,y,z$ but maintain the diagonal form. This permutation freedom 
will define my flavor symmetry, and we label one particular choice by the 
flavors, i.e.,
\be
S(q\bar q)
=\left( \begin{array}{ccc} 
S(u\bar u)& 0         &0 \\ 
0         & S(d\bar d)&0 \\
0         & 0         &S(s\bar s)
        \end{array}
\right ) \ee 
where the diagonal terms need not be equal once flavor symmetry is broken.

Of course, still under global color transformations of 
\underline{both} valence 
and sea one remains within the same hadron, - a $\pi^-$ remains a 
$\pi^-$ and a $K^-$ remains a $K^-$ etc. More precisely the charge and 
strangeness operators,
\be
Q=\left( \begin{array}{ccc} 
\frac 2 3& 0         &0 \\ 
0         & -\frac 1 3&0 \\
0         & 0         &-\frac 1 3
        \end{array} \right) , \ \ 
S=\left( \begin{array}{ccc} 
0& 0         &0 \\ 
0         & 0&0 \\
0         & 0&1
        \end{array}
\right) 
\ee 
which define\footnote{We remind the reader that flavor must be represented by 
complex or non-Hermitian fields, e.g., $|\pi^+>=(\lambda_1-i\lambda_2)/2$, 
and that charge and strangeness are a kind of interference term between the 
Hermitian and the anti-Hermitian part. In practice, in a C-invariant theory 
one can often neglect the anti-Hermitian part and consider instead flavorless,
Hermitian superpositions like $|\pi^+>+|\pi^->=\lambda_1$.}
 the charge as $q=\Tr[\Sigma Q \Sigma^\dagger-\Sigma^\dagger Q \Sigma] $
and strangeness $s=\Tr[\Sigma S \Sigma^\dagger-\Sigma^\dagger S \Sigma]$ 
 of a nonet meson $\Sigma $, must transform covariantly under color 
\be
Q'=UQU^\dagger,\ S'=USU^\dagger, \ \Sigma' =U\Sigma U^\dagger
\ee
which of course only implies that our choice of $u,d,s$ 
above is always done in a
particular, but arbitrary color reference frame, 
as chosen above for convenience

We can write for e.g. the vector flavor nonet when unmixed
\be
V=\left| \begin{array}{ccc} 
(\omega+\rho)/\sqrt 2 &      \rho^+         & K^{*+}\\ 
\rho^-               &(\omega-\rho)/\sqrt 2& K^{*0} \\
K^{*-}                  & \bar K^{*0}            & \phi
        \end{array} \right| \otimes  
\left| \begin{array}{ccc} 
S(u\bar u)& 0         &0 \\ 
0         & S(d\bar d)&0 \\
0         & 0         &S(s\bar s)
        \end{array}
\right>
\ee
Each meson is also a nonet in color (before the gluon dressing), 
i.e., we have 9 nonets in all. Flavor is a relative quantum number given 
by the ordering of $u,d,s$ in the valence part with respect to the ordering 
of the diagonal terms in the sea. Thus one transforms a $\pi^- $
to a $K^-$ by permuting $d \to s $ in the valence part but not in the sea.
On the other hand in a global color transformation one preforms 
an SU3 rotation 
(or a permutation of quark labels) in \underline{both} 
the valence and the sea part
of the wave function. 
In the limit of a symmetric sea, $S(u\bar u)=S(d\bar d)=S(s\bar s)$,
 all 9 degenerate nonets lie on top of each others, and both flavor 
and color is unbroken. But if the sea is asymmetric the flavor nonets are 
generally\footnote{A tricky point is that the sea can be somewhat asymmetric 
(since ${\bf 8\otimes 8}$ contains an octet), but still one can have a flavor 
symmetric spectrum. 
Thus e.g. all of the $d/u$ asymmetry seen in deep inelastic 
scattering on the proton need not result in isospin breaking. 
Only if the $d/u$ 
asymmetry in the proton is not equal to the $u/d$ asymmetry in the neutron do 
we have isospin violation. For the $s/d$ quark asymmetry the flavor symmetry 
breaking is more obvious if one knows that
in the sea of \underline{all} hadrons the $s$ quark is 
less frequent than the $d$.}
split, but color remains always exact.

A natural mechanism for generating the asymmetric 
sea is given by quantum loops 
such as $K^*\to K^*\pi,K\pi,K\phi, ... \to K^*$ or 
$\pi\to \pi \sigma, K^*\bar K, ... \to \pi$ etc.
Hadrons are, in fact, unique compared to other bound states, 
like atoms or nuclei, 
in that they are partly composed of the hadrons themselves, 
although the latter 
are in virtual off shell states. A proton is part of the time a proton and a pion, 
and a pion is part of the time in a three pion state etc. Constituent quarks are 
again composed of virtual quarks and gluons
\be
\small{
|q> \propto |q>(1+\alpha|q\bar q> +\beta |q\bar qq\bar q>+ ...)\times gluons } \label{quark}
\ee
i.e., the same constituents occur on both the l.h.s. and the r.h.s. 
Introducing a reference frame
for the quarks, the same reference frame appear on both sides of the equation. 
A color rotation is like rotating an object (valence) \underline{and} the observer (sea) 
resulting in no change for the observer.
On the other hand a change (permutation) in part of a state (the valence part) while 
the rest remains intact results in a new (flavor) state.

Since a strange-antistrange  ($K^+ K^-$) virtual state should be less frequent than 
a non strange ($\pi^+\pi^-$) virtual state loops naturally lead to an asymmetric 
sea in all hadrons, where
\ss\ is less frequent than \uu\ or \dd . Furthermore, this mechanism can be self-enhancing
resulting in an instability: A small initial strange-nonstrange splitting for the (input)
virtual states generates a bigger splitting in the output physical state.

Although our ansatz that flavor symmetry is broken by an asymmetric sea, rather than by the
infinite vacuum might seem to be a minor one, it has dramatic consequences.

The most important one is that the symmetry can be broken spontaneously, without the
appearance of massless Goldstone bosons. This has been a major stumbling block in 
previous attempts to break flavor symmetry spontaneously. The  discrete flavor symmetry 
introduced above can thus be broken spontaneously without the necessary appearance of 
unseen Goldstone bosons,  and the continuous SU3 symmetry is never broken. 
In a series of previous publications\cite{NAT} actual dynamic calculations
of such spontaneous symmetry breaking was preformed. These involved only 
scalars in a simplified model, and pseudoscalars and vectors in a little more 
realistic model. It was demonstrated that
the symmetry breaking goes in the right direction compared to experiment, 
and that e.g. the mechanism naturally explains the approximate nature of the 
Okubo-Zweig-Iizuka rule, the near ideal mixing and the equal spacing rule of meson 
multiplets. 

Furthermore this opens up a new scenario for predicting quark masses (and 
possibly the CKM matrix), since one can start from an exactly symmetric theory with 
few parameters (and in accord with the flavor blindness of QCD) which can have both 
an unstable symmetric and a stable flavor asymmetric solution. Then, of course, only 
the stable asymmetric one is the true physical solution.

Of course our discrete symmetry can also be broken explicitly. 
In fact, one expects that
electro-weak interactions should break the up-down 
symmetry and give the $u,c,t$ quarks
an extra mass, while the $d,s,b$ quarks may be degenerate or nearly 
massless before the spontaneous breaking by strong interactions. 
A true lepton-quark symmetry might emerge. 

There is some similarity with the color-flavor connection discussed here,
and the color-flavor locking of Sch\"afer and Wilczek\cite{schafer} and 
collaborators, although the latter is applied within another context of 
high density QCD, where the \qq\ sea and the vacuum merge.

The suggested new interpretation of flavor symmetry also throws some new 
light on the nature of superselection rules\cite{www}, which was vigorously 
discussed almost half a century ago when isospin invariance had been introduced.

\end{document}